\documentclass[preprint2]{aastex}

\shorttitle{BLAPs: the possible surviving companions of SNe Ia}
\shortauthors{Meng, Han, Podsiadlowski \& Li}

\begin{document}


\title{Blue Large-Amplitude Pulsators: the Possible Surviving Companions of Type Ia Supernovae}


\author{Xiang-Cun Meng$^{\rm 1,2,3}$, Zhan-Wen Han$^{\rm 1,2,3}$, Philipp Podsiadlowski$^{\rm 4}$, Jiao Li$^{\rm 5,1}$}
\affil{$^{\rm 1}$Yunnan Observatories, Chinese Academy of Sciences, 650216 Kunming, PR China\\
$^{\rm 2}$ Key Laboratory for the Structure and Evolution of
Celestial Objects, Chinese Academy of Sciences, 650216 Kunming, PR
China\\
$^{3}$Center for Astronomical Mega-Science, Chinese Academy of
Sciences, 20A Datun Road, Chaoyang District, Beijing, 100012, PR China\\
$^{\rm 4}$Department of Astronomy, Oxford University, Oxford OX1
3RH, UK\\
$^{\rm 5}$Key Laboratory of Space Astronomy and Technology,
National Astronomical Observatories, Chinese Academy of Sciences,
Beijing 100101, PR China} \email{xiangcunmeng@ynao.ac.cn}





\begin{abstract}
  The single degenerate (SD) model, one of the leading models for the
  progenitors of Type Ia supernovae (SNe Ia), predicts that there
  should be binary companions that survive the supernova explosion
  which, in principle, should be detectable in the Galaxy.  The
  discovery of such surviving companions could therefore provide
  conclusive support for the SD model. Several years ago, a new type
  of mysterious variables was discovered, the so-called blue
  large-amplitude pulsators (BLAPs). Here we show that all the
  properties of BLAPs can be reasonably well reproduced if they are
  indeed such surviving companions, in contrast to other proposed
  channels. This suggests that BLAPs could potentially be the
  long-sought surviving companions of SNe Ia.  Our model also predicts
  a new channel for forming single hot subdwarf stars, consistent with
  a small group in the present hot-subdwarf-star sample.
\end{abstract}





\section{INTRODUCTION}\label{sect:1}
The nature of the progenitors of Type Ia supernovae (SNe Ia) remains a
hotly debated topic (\citealt{HN00}; \citealt{WANGB12};
\citealt{MAOZ14}), even though they have been so important for
determining cosmological parameters(\citealt{RIE98}; \citealt{PER99};
\citealt{MENGXC15}). At present, a basic framework has been
established where a SN Ia originates from the thermonuclear explosion
of a carbon-oxygen white dwarf (CO WD) in a binary system
(\citealt{HF60}). The WD accretes material from its companion and
increases its mass close to its maximum stable mass, where a
thermonuclear explosion occurs in the WD (\citealt{BRA04}).  Based on
the nature of the companion star of the accreting WD, two classes of
progenitor scenarios have been proposed: the single degenerate (SD) model where
the companion is a non-degenerate star, i.e.\ a main-sequence or a slightly
evolved star (WD+MS), a red giant star (WD+RG) or a helium star (WD +
He star) (\citealt{WI73}; \citealt{NTY84}), and the double-degenerate
(DD) model involving the merger of two CO WDs (\citealt{IT84};
\citealt{WEB84}). Both models have some support on both the
observational and the theoretical side (\citealt{HOWELL11}).

A basic difference between the two classes of models is that there
still is a surviving companion after the supernova explosion in
the SD model but not in the DD model (but see \citealt{SHENK18}).
Searching for surviving companions directly in supernova remnants
(SNRs) is a viable way to distinguish between the different
models. The discovery of potential surviving companions in some
supernova remnants has revealed the power of the method
(\citealt{RUIZLAPUENTE04}; \citealt{LIC17}). However, the typical
lifetime of a SNR is only a few $10^{\rm 4}$ yr
(\citealt{SARBADHICARY17}), which is much shorter than the
lifetime of any surviving companion. Therefore, there must be a
large number of surviving companions in the Galaxy which are not
associated with SNRs, freely cruising in space, if the SD model
contributes, at least in part, to the production of SNe Ia. The
surviving companion may show some unusual properties compared to
normal single stars, e.g.\ an atmosphere polluted by supernova
ejecta and a relative high space velocity (\citealt{HAN08}). If
such surviving companions were discovered, this could provide
conclusive support for the SD model.

Recently, \citet{PIETRUKOWICZ17} found a new class of variable
stars named blue large-amplitude pulsators (BLAPs), objects whose
origin is still a complete mystery. BLAPs are single,
hydrogen-deficient stars associated with the Galactic disc, and no
BLAP has been discovered in the Magellanic Clouds
(\citealt{PIETRUKOWICZ18}); i.e.\ BLAPs appear to belong to a
young population with a relatively high metallicity. Model
simulations show that BLAPs are core-helium-burning or
shell-hydrogen-burning stars, and that their total mass is smaller
than $\sim 1.2\, M_{\odot}$ (\citealt{PIETRUKOWICZ17}). Their
positions in the Hertzsprung-Russell (HR) diagram locate them
between main-sequence stars and hot sdOB stars, as do their
surface gravities; this suggests that the envelopes of BLAPs are
slightly more massive than those of hot sdOB stars. However, the
lifetime of the shell-hydrogen-burning stars during the BLAP stage
is too short compared with the lifetime of the BLAPs deduced from
the rate of their period change (\citealt{PODSIADLOWSKI02};
\citealt{WUT18}; \citealt{ROMERO18}; \citealt{BYRNE18};
\citealt{CORSICO18}). This leaves core-helium-burning stars with a
thin hydrogen envelope as the only viable solution for BLAPs
(\citealt{WUT18}; \citealt{BYRNE18}).  To form such a special
structure, a star needs to lose its hydrogen-rich envelope after a
helium core has formed in its center.

\begin{figure*}
\centerline{\includegraphics[angle=0,scale=.60]{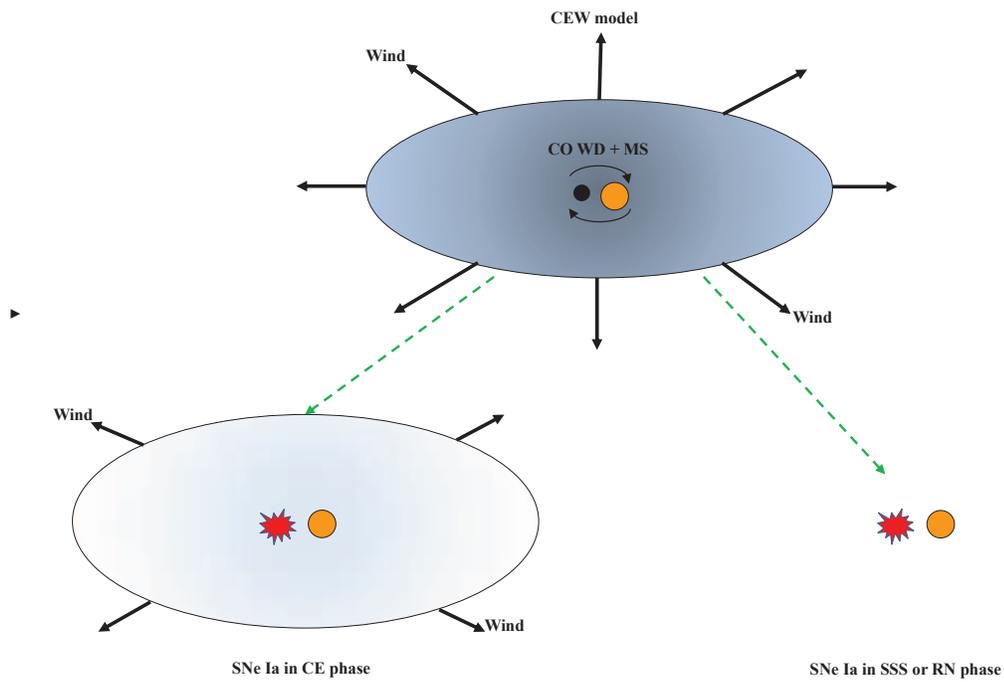}}
\caption{Schematic diagram illustrating the common-envelope wind
(CEW) model, where the SN Ia may explode in a CE, SSS or RN phase.
BLAPs originate from those exploding in the CE
phase.}\label{bb}
\end{figure*}

\begin{figure}
\centerline{\includegraphics[angle=270,scale=.35]{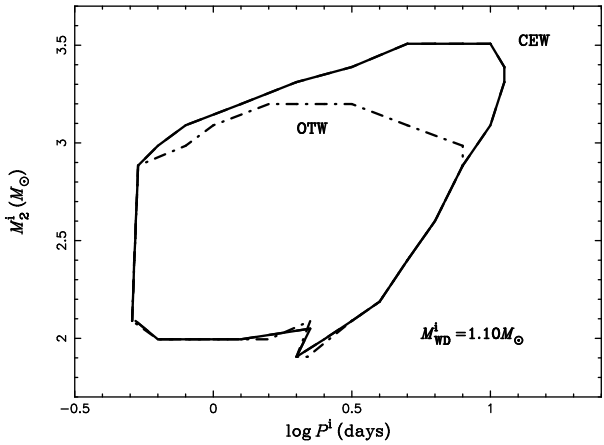}}
\caption{Comparison between the CEW and the OTW model for the initial
parameter contours leading to SNe Ia, where the initial WD mass is
1.10\,$M_{\odot}$. The data for the plot are taken from \citet{MENG09}
and \citet{MENGXC17a}.}\label{compare}
\end{figure}

\citet{MENGXC17a} developed a new version of the SD model, which they
named the common-envelope wind (CEW) model. In this model, mass
transfer between a WD and its companion can begin when the companion
is a MS star or is crossing the Hertzsprung gap (HG). If the
mass-transfer rate exceeds the critical accretion rate of the WD, the
WD will expand to a RG-like object, and a common envelope (CE) is
assumed to form around the binary system. The WD then gradually
increases its mass at the base of the CE. For a low density of the CE,
the binary system is expected to survive from the CE phase until the
WD approaches the Chandrasekhar mass and explodes as a SN Ia.  The WD
may explode while it is still in the CE phase, a phase of stable
hydrogen burning (and appear as a supersoft X-ray sources [SSS]), or a
phase of weakly unstable hydrogen burning, where the system would
appear as a recurrent nova (RN), as illustrated in
Fig.~\ref{bb}. According to the different phases when the SN Ia occurs
in the CEW model, even some peculiar SNe Ia may share the same origin:
e.g.\ the so-called SNe Ia-CSM and 02cx-like objects may both
originate from the explosions of hybrid carbon-oxygen-neon (CONe) WDs
in SD systems (\citealt{MENGXC17b}).  In the CEW model, if the mass
transfer for a binary system begins when the companion crosses the HG,
a helium core has been formed in the center of the companion. The core
mass is determined by the initial companion mass and the initial
orbital period: the more massive the initial companion or the longer
the initial orbital period, the more massive the helium core of the
companion.  After the supernova explosion, such a companion may become
a hydrogen-deficient low-mass single star (e.g.\ Fig. 19 in
\citealt{MENGXC17a}, where the SN Ia explodes in the CE phase), and
share many properties with BLAPs, making them promising candidates for
surviving companions from SD systems. Here, adopting the CEW model, we
will show that all the properties of BLAPs may be simultaneously
reproduced by the surviving companions in the CEW model: their
population characteristics, their single-star nature, their lifetime
as a BLAP, their location in the HR diagram, their surface helium
abundance and surface gravity, their radial velocity, their pulsation
periods, including the rate and sign of the period change, the total
number of BLAPs and the number ratio of BLAPs to hot subdwarf stars in
the Galaxy. We will also show that there is no other proposed channel
that can simultaneously explain all these properties. Hence we suggest
that BLAPs are likely surviving companions of SNe Ia, as predicted by
the SD model, and that their discovery provides strong evidence in
support of the SD model.

In section \ref{sect:2}, we describe our methods and the main
results of our calculations. In section \ref{sect:3} we discuss
the results, and present our main conclusions in section
\ref{sect:4}.

\begin{figure}
\centerline{\includegraphics[angle=270,scale=.35]{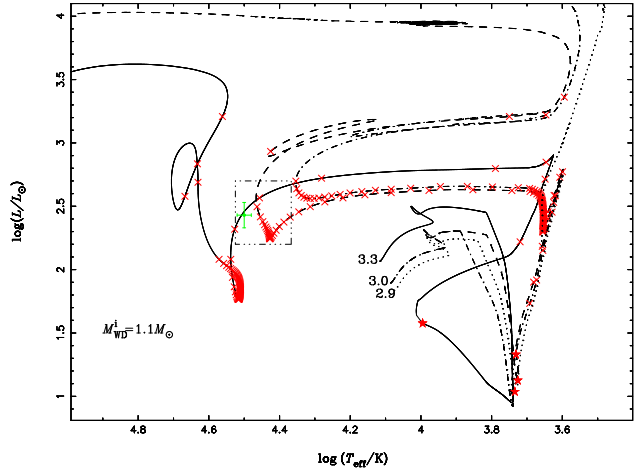}}
\caption{The evolutionary tracks of the companion stars in the HR
  diagram.  Red stars show the position where the SNe Ia are
  assumed to occur, and the dash-triple-dotted rectangle presents the
  region for BLAPs. The initial WD masses are the same for the four
  systems, i.e.\ $M_{\rm WD}^{\rm i}=1.1\,M_{\odot}$. The initial
  companion masses and periods of the four systems are [$M_{\rm 2}^{\rm i}/M_{\odot}$, $\log(P^{\rm i}/{\rm d})$] = (3.3, 0.9),
  (3.0, 0.8), (3.0, 0.6) and (2.9, 0.7), and the evolutionary tracks
  of the companions from the four systems are shown by solid, dashed,
  dash-dotted and dotted curves, respectively. The age interval
  between adjacent crosses is $10^{\rm 6}$ yr. The green cross
  represents BLAP-009, whose luminosity is calculated based on the
  distance from the GAIA DR2 and the average apparent magnitude in
  \citet{PIETRUKOWICZ17}, where the error bar of the luminosity is
  determined from the distance error in GAIA DR2, and the error bar of
  the effective temperature comes from the spectral fitting in
  \citet{PIETRUKOWICZ17}.}\label{hrd}
\end{figure}

\section{METHODS AND RESULTS}
\label{sect:2}




\begin{figure}
\centerline{\includegraphics[angle=270,scale=.35]{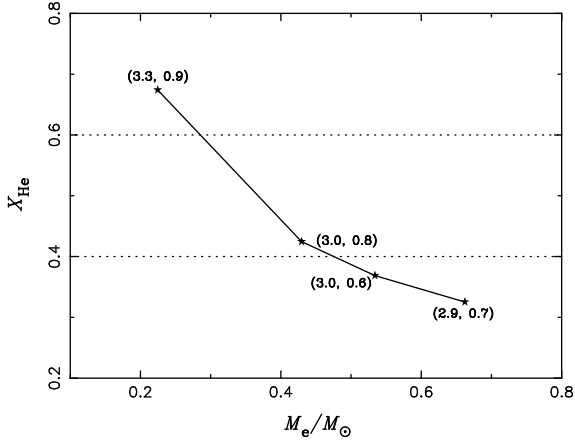}}
\caption{Correlation between the surface helium abundance and the
  envelope mass of the companion when $M_{\rm WD}=1.378~M_{\odot}$
  (solid curve), where the four points correspond to the four systems
  in Fig.~\ref{hrd}. The dotted curves show the likely regions of the
  surface helium abundance for BLAPs when $M_{\rm WD}=1.378~M_{\odot}$
  (see section~\ref{sect:2.3}).}\label{mehe}
\end{figure}

\subsection{HR diagram}\label{sect:2.1}
In the CEW model, if mass transfer between a relatively massive
initial WD and a relatively massive initial companion begins in the
HG, the companion can become a hydrogen-deficient, low-mass single
star after the supernova explosion (e.g.\ Fig. 19 in
\citealt{MENGXC17a}). The surviving companion then has a helium core
and a thin hydrogen-rich envelope. If helium is ignited in the core,
the star will become a core-helium-burning star with a thin
hydrogen-deficient envelope. As far as the companion properties after
the supernova explosion are concerned, the difference for most cases
between the CEW and the optically thick wind (OTW) model is not very
significant (\citealt{HAC96}; \citealt{MENGXC17a}). However, as shown
in \citet{MENGXC17a}, some systems that cannot produce SNe Ia in the
OTW model may do so in the CEW model (the upper-right region in the
$P^{\rm i}-M_{\rm 2}^{\rm i}$ plane in Fig.~\ref{compare}). Indeed, it
is just these systems that are more likely to leave
hydrogen-deficient, single-star companions.  Considering the merits of
the CEW model relative to the OTW model, we here use the CEW model to
calculate the evolution of the companion.

We assume that the WDs explode as SNe Ia when $M_{\rm
WD}=1.378~M_{\odot}$. Here, we do not consider the effects of
spin-up/spin-down and stripping-off on the companions since there
are still many uncertainties on how to implement these effect, but
we note that these are unlikely to change our basic conclusions
(see the discussions in Section. \ref{sect:3.2}). After the
supernova explosions, the companions may become
hydrogen-deficient, single stars, such as BLAPs. To examine whether
the surviving companions can reproduce the other properties of
BLAPs, we choose four typical binary systems and continue to
evolve the companion stars after the supernova explosion and
record their various parameters that may directly be compared with
the properties of BLAPs.

In Fig.~\ref{hrd}, we show the evolutionary tracks of the
companions in the HR diagram. Generally, the companions ascend the
red-giant branch (RGB) after the supernova explosion, and helium
is ignited in the core at the tip of the RGB. The companions then
become horizontal-branch (HB) stars and stay on the HB, while
shell hydrogen burning above the helium-burning core continues to
consume hydrogen-rich envelope material.  However, depending on
the different envelope masses of the companions on the HB, their
subsequent evolution can become quite different. If the envelope
of the companion is so thick that it cannot be consumed completely
before the exhaustion of the helium in the center, the star will
evolve like a typical asymptotic-giant-branch (AGB) star (dotted
line). In contrast, if the envelope is so thin that it is
exhausted soon after helium ignition, the evolutionary track of
the companion is similar to a hot subdwarf star (solid line,
\citealt{HANZW02,HANZW03}). For the companion from the system with
initial parameters of [$M_{\rm WD}^{\rm i}/M_{\odot}$, $M_{\rm
2}^{\rm i}/M_{\odot}$, $\log(P^{\rm i}/{\rm d})$] = (1.1, 3.0,
0.8), the envelope is neither very thin nor very thick. As the
envelope is consumed due to shell hydrogen burning on the HB, the
envelope becomes thinner and thinner and the effective temperature
increases correspondingly. As a result, the evolutionary track of
the companion moves to the left and may cross the region where
BLAPs are located in the HR diagram (dashed line). At the BLAP
stage, shell hydrogen burning is extinguished, but there is still
a very thin hydrogen-deficient envelope left as has been deduced
for BLAPs (\citealt{PIETRUKOWICZ17}). Since the companion has
spent a long time on the HB, the lifetime of the companion in the
BLAP stage is shorter than that of a typical hot subdwarf stars,
but can still be as long as a few $10^{\rm 7}$ yr as inferred for
BLAPs (\citealt{PIETRUKOWICZ17}). This suggests that BLAPs are in
the middle or late phase of helium core burning (see also
\citealt{WUT18}). In addition, the evolutionary track of the
companion from the system with [$M_{\rm WD}^{\rm i}/M_{\odot}$,
$M_{\rm 2}^{\rm i}/M_{\odot}$, $\log(P^{\rm i}/{\rm d})$] = (1.1,
3.0, 0.6) is close to the region of BLAPs, but with a somewhat
lower effective temperature because of its thicker envelope.

The different evolutionary tracks of the companions in
Fig.~\ref{hrd} are therefore mainly due to the different envelope
masses at the time of the supernova explosion. BLAPs are
hydrogen-deficient, which implies that their progenitors could
also be hydrogen-deficient when they were born. Based on the
results in \citet{MENGXC17a}, for a system where mass transfer
begins when the companion crosses the HG, the surface helium
abundance of the companion when $M_{\rm WD}=1.378~M_{\odot}$ has a
strong dependence on its envelope mass (e.g.\ the evolution of
$\dot{M}_{\rm WD}$ in Figs. 4 and 19 in \citealt{MENGXC17a}). In
Fig.~\ref{mehe}, following the definition of the core as in
\citet{HAN94} and \citet{MENGXC08}, we show the correlation
between the surface helium abundance and the envelope mass when
$M_{\rm WD}=1.378~M_{\odot}$, where the envelope mass, $M_{\rm
e}$, is defined as the difference between the companion mass,
$M_{\rm 2}^{\rm SN}$, and the core mass, $M_{\rm c}$. The figure
shows a clear anti-correlation between the envelope mass and the
surface helium abundance, as expected. Therefore, the surface
helium abundance may be taken as an indicator of the envelope mass
at the time of supernova explosion. Similarly, Fig.~\ref{hrd}
shows that, the lower the envelope mass at the time of the
explosion, the more the subsequent evolutionary track will
resemble the track of a hot subdwarf star (i.e.\ will become
hotter with lower envelope mass).

\begin{figure}
\centerline{\includegraphics[angle=270,scale=.35]{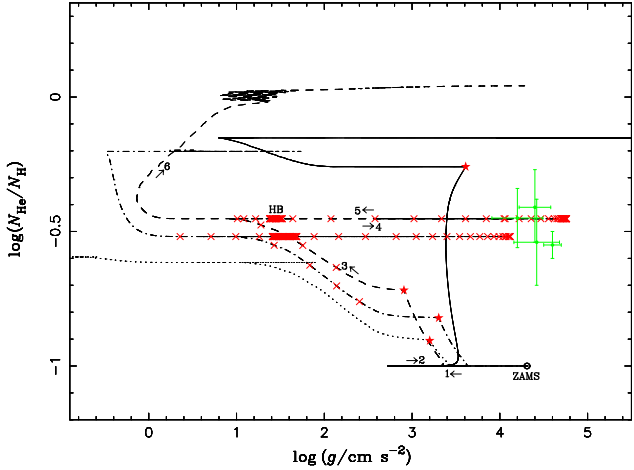}}
\caption{The evolution of the surface helium abundance and gravity of
  the surviving companions of SNe Ia. The lines represent the same
  systems as shown in Fig.~\ref{hrd}, and the red stars show the
  positions where supernova explosions are assumed to take place. The
  green crosses represent the four BLAPs with spectroscopic observations in
  \citet{PIETRUKOWICZ17}, and the age interval between adjacent
  crosses is $10^{\rm 6}$ yr. The arrow and numbers mark the
  evolutionary direction of the model with [$M_{\rm WD}^{\rm i}/M_{\odot}$, $M_{\rm 2}^{\rm i}/M_{\odot}$, $\log(P^{\rm i}/{\rm d})$] = (1.1, 3.0, 0.8) in the plot.}\label{gxhe}
\end{figure}

\begin{figure}
\centerline{\includegraphics[angle=270,scale=.35]{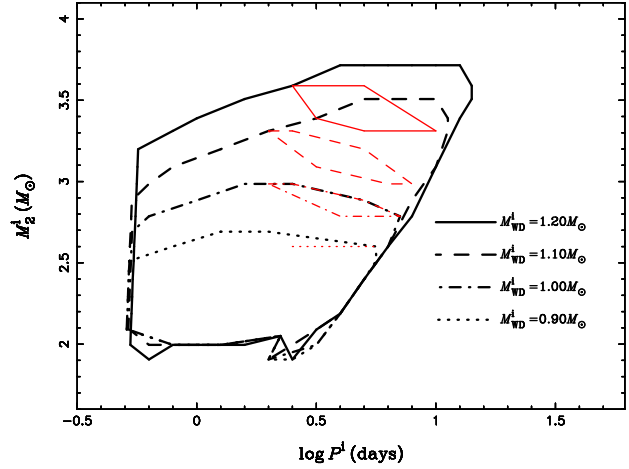}}
\caption{The initial parameter regions for different initial WD
masses for BLAPs (red thin lines). For comparison, the thick lines
present the parameter spaces for SNe Ia (reproduced from
\citealt{MENGXC17a}).}\label{gblap}
\end{figure}

\begin{figure}
\centerline{\includegraphics[angle=270,scale=.35]{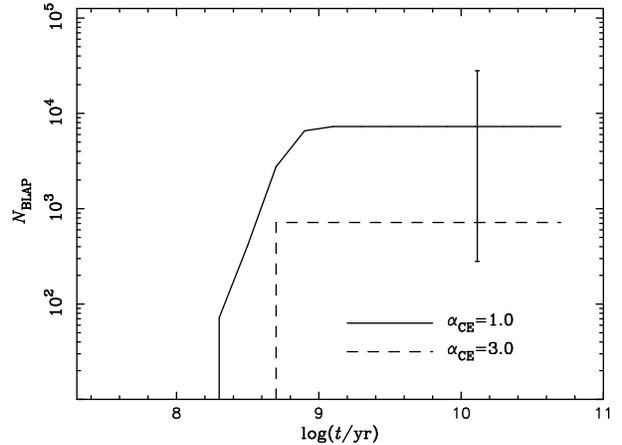}}
\caption{The evolution of the number of BLAPs in the Galaxy, where a
constant star formation rate of $5\,M_{\odot}$/yr is assumed. The
vertical bar shows the estimated region of the number of BLAPs in
the Galaxy.}\label{nblap}
\end{figure}

\subsection{The helium abundance and gravity of BLAPS}\label{sect:2.2}
As \citet{PIETRUKOWICZ17} showed, BLAPs are helium-rich, and their
surface gravities lie between main-sequence stars and the known
sdOB stars. If BLAPs are the surviving companions of SNe Ia, the
companion predicted by the SD model will reproduce their surface
helium abundance and surface gravity simultaneously. In
Fig.~\ref{gxhe}, we show the evolution of the surface helium
abundance and surface gravity of the surviving companions, where
the initial systems are the same as those in Fig.~\ref{hrd}. These
figures show that, after the supernova explosion, the surviving
companion from the system with [$M_{\rm WD}^{\rm i}/M_{\odot}$,
$M_{\rm 2}^{\rm i}/M_{\odot}$, $\log(P^{\rm i}/{\rm d})$] = (1.1,
3.0, 0.8) experiences a phase where the surface gravity decreases
while the surface helium abundance increases until the star
arrives on the HB. In this phase, the companion ascends the RGB,
where it experiences the first dredge-up, which leads to the
mixing up of helium-rich material and an increase of the surface
helium abundance. At the same time, the expansion of the star
reduces the surface gravity. After the companion has settled on
the HB, large-scale convection ceases in the envelope, and the
surface helium abundance no longer changes. With the consumption
of the envelope on the HB, the radius of the companion decreases,
and the surface gravity increases until the companion becomes a
BLAP (based on its position in the HRD; Fig.~\ref{hrd}). This
demonstrates that our surviving-companion model can simultaneously
reproduce the surface helium abundance and the gravity of the
BLAPs observed in \citet{PIETRUKOWICZ17}. In particular, the
surface gravity of the surviving companion in the BLAP phase lies
between MS and hot subdwarf stars. For the star with the thinnest
envelope at the time of the supernova explosion, the surface
gravity during the core-helium-burning phase is higher than that
of BLAPs but is consistent with sdOB stars, and the surface helium
abundance and surface gravity of the companion from the system
with [$M_{\rm WD}^{\rm i}/M_{\odot}$, $M_{\rm 2}^{\rm
i}/M_{\odot}$, $\log(P^{\rm i}/{\rm d})$] = (1.1, 3.0, 0.6) are
also close to those of BLAPs, as shown in Fig.~\ref{hrd}.

\begin{figure}
\centerline{\includegraphics[angle=270,scale=.35]{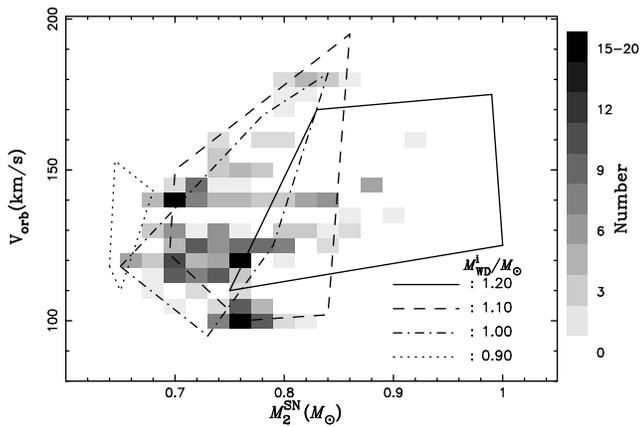}}
\caption{The distributions of companion mass and orbital velocity
when $M_{\rm WD}=1.378~M_{\odot}$ for a constant star-formation
rate and $\alpha_{\rm CE}=1.0$. The lines present the final
parameter space of the companion mass and orbital velocity for
different initial WD masses.}\label{blap}
\end{figure}

\subsection{The number of BLAPs in the Galaxy}\label{sect:2.3}
The Optical Gravitational Lensing Experiment (OGLE) project
surveyed about 5\% of the Milky Way disc and found 14
BLAPs (\citealt{PIETRUKOWICZ17}, private communication). At
present, it is very difficult to estimate the completeness of the
sample, and some BLAPs must be hidden behind clouds of dust in the
surveyed directions. Here, as a very conservative upper limit, we
assume that the number of BLAPs missed could be as high as 99\%;
this would give an estimate for the number of BLAPs in the Galaxy
roughly between 280 and 28000.


Fig.~\ref{hrd} shows that not all surviving companions have properties
consistent with those of BLAPs. To estimate the number of BLAPs from
the SD model, we need to know the initial parameter space producing
them. Here, we do not recalculate grids of binary evolution sequences
to determine this parameter space; instead we just try to constrain it
from the model grids already calculated in \citet{MENGXC17a}, adopting
some additional constraints. From Figs.~\ref{hrd} and \ref{mehe} we
know that the envelope mass of the companion at the time of the
supernova explosion is the key parameter determining whether the
surviving companion produces the properties of BLAPs, and the envelope
mass is anti-correlated with the surface helium abundance. Stars with
a much higher or a much lower surface helium abundance produce
envelopes that are either too thin or too thick and do not reproduce
the location of BLAPs in the HR diagram. As we will discuss in Section
\ref{sect:2.5}, if a star is not located in this region, it will
probably not show the pulsation modes of BLAPs because of the
different surface gravity or different mean density.  In addition, the
binary evolution calculations in this paper do not support a companion
star with a mass $>1.1~M_{\odot}$ at the time of the supernova
explosion as a progenitor of a BLAP (the dotted line in
Figs.~\ref{hrd}, see also the model simulations in
\citealt{PIETRUKOWICZ17}). Considering that the observed region of
BLAPs in the HR diagram is mainly constrained by a single BLAP and
that the real region could be much larger, we here assume somewhat
arbitrarily that the helium abundance has to be between 0.4 and 0.6
and the companion mass less than 1.1 $M_{\odot}$ at the time of the
explosion, so that the surviving companion can become a BLAP after
central helium ignition. With these constraints, we can use the grids
in \citet{MENGXC17a} to determine the parameter space that produces
BLAPs; this is shown in Fig.~\ref{gblap}. This clearly shows that the
initial systems that produce BLAPs consist of relative massive WDs
with massive companions and have a relative long initial period,
i.e.\ mass transfer begins when the companion crosses the HG. This
parameter space is only a small part of the whole parameter space that
leads to SNe Ia (see Fig.~\ref{gblap}), implying that the birth rate
of BLAPs, $\nu$, is much lower than the overall rate of SNe Ia from
the SD model.

Based on the above parameter space, we performed two binary
population synthesis (BPS) simulations using the rapid binary
evolution code developed by \citet{HUR00,HUR02}, where the BPS
method is the same as described in \citet{MENGXC17a}. For the BPS
simulations, the common-envelope ejection efficiency, $\alpha_{\rm
CE}$, is the key parameter affecting the birth rate of BLAPs.
Following \citet{MENGXC17a}, we take $\alpha_{\rm CE}=1.0$ or
$\alpha_{\rm CE}=3.0$. As the parameter space producing BLAPs is
so much smaller than that producing SNe Ia, our calculations show
that only 0.3\% to 3.3\% of all SNe Ia produce BLAPS.  As shown in
Fig.~\ref{hrd}, the lifetime of a BLAP is shorter than that of a
typical hot subdwarf star since the progenitor of the BLAP spends
part of its life in the HB phase. Simply assuming that all BLAPs
have a lifetime of $\tau=5\times10^{\rm 7}$ yr, we may obtain the
evolution of the number of BLAPs with time in the Galaxy by
$\nu\times \tau$, as shown in Fig.~\ref{nblap}. The predicted
number of BLAPs is roughly between 750 and 7500, very much
consistent with the rough estimate of BLAPs made earlier.

\begin{figure}
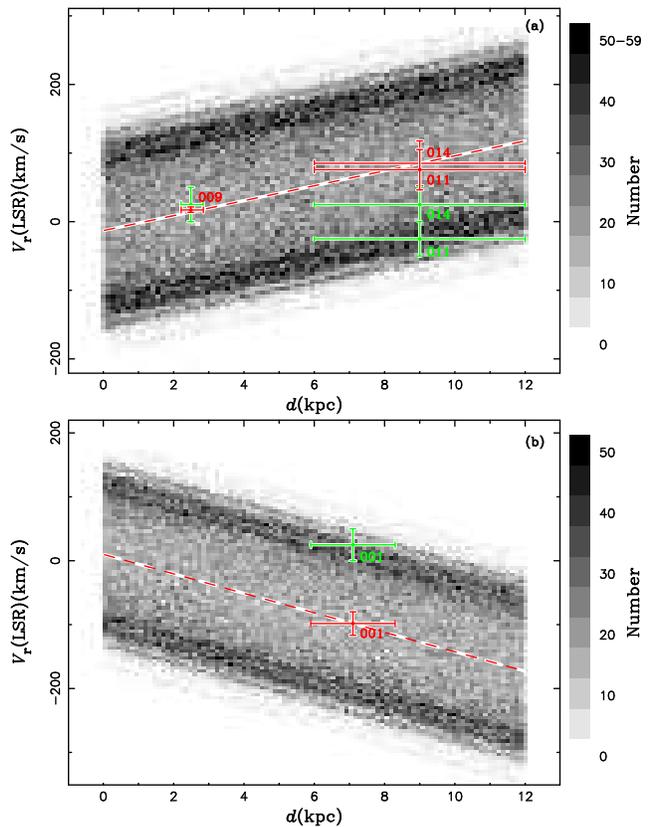

\centerline{\includegraphics[angle=270,scale=.35]{rvblap1.ps}}
\centerline{\includegraphics[angle=270,scale=.35]{rvblap2.ps}}
\caption{The distribution of the radial velocity in the local standard
of rest (LSR) frame versus distance for BLAPs, for the case $\alpha_{\rm CE}=1.0$.
The red crosses show the positions if the observed BLAPs are disc
stars, while the green crosses show the measured values
(\citealt{PIETRUKOWICZ17}, private communication). The dashed
lines show the correlation between radial velocity and distance in
the LSR frame for disc stars, obtained from equation (\ref{vrd})
for the directions of BLAP-014 [panel (a)] and BLAP-001 [panel
(b)].}\label{rvblap}
\end{figure}

\subsection{The distribution of radial velocity}\label{sect:2.4}
At the time of the supernova explosion, the companions in close
binaries have relatively large orbital velocities, and the
companions will inherit this orbital velocity as runaway space
velocity after the WD has been disrupted. Fig.~\ref{blap} shows
the distributions of the orbital velocity and companion mass when
$M_{\rm WD}=1.378~M_{\odot}$ for the systems where the companions
will become BLAPs.  Most of the companions have a mass of
$0.76\pm0.1~M_{\odot}$ and all companions have masses below
$1\,M_{\odot}$ although our formal adopted constraint was less
than 1.1 $M_{\odot}$. This mass range is very much consistent with
theoretical pulsation model constraints ($\sim0.7-1.1$
$M_{\odot}$, \citealt{WUT18}). In addition, the companion stars
have space velocities between 100 km/s and 200 km/s relative to
the centre of the mass of the binary systems. Such a high space
velocity should be reflected in the radial velocities of BLAPs.

To obtain the distribution of the radial velocity of the predicted
BLAPs, we performed a Monte Carlo simulation, where the radial velocity of a
surviving companion is determined by
 \begin{equation}
V_{\rm r}=V_{\rm orb}\cdot\cos i + V_{\rm r, disc},\label{vr}
  \end{equation}
where $i$ is the angle between the space velocity and the line of
sight and $i$ is generated randomly. $V_{\rm r, disc}$ is the
radial velocity relative to the local standard of rest (LSR) for a
disc star at a Galactic longitude, $l$, and a distance, $r$, and
is determined by
 \begin{equation}
 V_{\rm r, disc}=-V_{\odot}\cdot\cos(l-l_{\odot}) + A\cdot r\cdot\sin(2l),\label{vrd}
  \end{equation}
where $l$ and $l_{\odot}$ are the Galactic longitudes of a disc star
and the solar apex, respectively, $r$ is the distance of the star,
$V_{\odot}$ is the Sun's velocity in the LSR, and $A$ is Oort's constant
(\citealt{BOVY17}). Here, $r$ is also generated in a Monte Carlo way,
while $l$ is taken to be towards the directions of BLAP-014 [panel (a)
  in Fig.~\ref{rvblap}] and BLAP-001 [panel (b) in Fig.~\ref{rvblap}].
In this discussion, we do not consider any Galactic dynamics, since
the surviving companions may only travel about $\sim2$ kpc in the
Galaxy before they become BLAPs, and their positions and velocities are
therefore not significantly affected by dynamical effects.

Fig.~\ref{rvblap} shows the distribution of the radial velocity of
the surviving companions versus distance.  The radial velocity has
a larger scatter due to the different orbital velocities and
inclination angle $i$. The distribution of the radial velocity at
a given distance $r$ has two peaks at a velocity of about $V_{\rm
r, disc}\pm110~{\rm km/s}$, which is mainly caused by the
different orbital velocities of the surviving companion when
$M_{\rm WD}=1.378~M_{\odot}$. The correlation between $V_{\rm r,
disc}$ and $r$ is shown by two dashed lines, where one corresponds
to the direction of BLAP-001 and the other is for the direction of
BLAP-014. In the figure, we also plot the radial velocity of the
BLAPs with spectral observations, where the red crosses assume
that BLAPs are normal disc stars, and green crosses show the
observed values (\citealt{PIETRUKOWICZ17}, private communication).
Three of the four BLAPs have quite different radial velocities
from disc stars in their directions and at their distances.
Especially, BLAP-001 has a positive radial velocity but should be
negative if BLAPs-001 were a disc star, while BLAP-011 has a
negative radial velocity but should be positive if BLAP-011 were a
disc star. The radial velocity differences between the BLAPs and
the disc stars at the same position are as high as $123\pm45$
km/s. Generally, at a given distance and in a given direction, the
scatter in the radial velocity of disc stars should be less than
$\sim20$ km/s (\citealt{DEHNEN98}; \citealt{ANGUIANO18}). So, the
difference of the radial velocity between the BLAPs and the normal
disc stars cannot be simply explained by the scatter of the radial
velocity of the disc stars, and other mechanisms are required to
explain the difference. Interestingly, the observed values of the
radial velocity for BLAPs-001, 011 and 014 are located around the
peak region in Fig.~\ref{rvblap}. Hence, the orbital velocity
could provide a reasonable explanation for the difference.

\begin{figure}
\centerline{\includegraphics[angle=270,scale=.35]{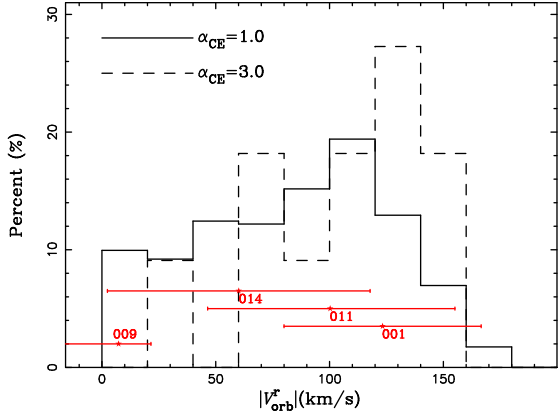}}
\caption{The distribution of the radial component of the orbital
velocity of the companion when $M_{\rm WD}=1.378~M_{\odot}$ for
different $\alpha_{\rm CE}$, where a constant star-formation rate
is assumed. The horizontal bars show the difference of the radial
velocity between the observed values and the disc stars for
the BLAPs with spectral observations.}\label{rvdis}
\end{figure}

If the difference of the radial velocity between the BLAPs and the
disc stars mainly originates from the orbital velocity of the
companion at the moment of supernova explosion, we would expect that
the radial component of the orbital velocity of the companion could
reproduce the difference. Here, the radial component of the orbital
velocity, $|V_{\rm orb}^{\rm r}|$, is set to be $|V_{\rm orb}\cdot\cos
i|$, where $i$ is again generated in a Monte Carlo way. In Fig.~\ref{rvdis},
we show the distribution of the radial component of the orbital
velocity of the companions for different $\alpha_{\rm CE}$ and also
the difference of the radial velocity between the observed values and
the disc stars for four BLAPs.  There is a peak in the distribution,
irrespective of the value of $\alpha_{\rm CE}$, consistent with those
in Fig.~\ref{rvblap}.  Fig.~\ref{rvdis} shows that most surviving
companions ($\sim$70\% to 80\%) have a radial velocity component
between 50 km/s and 150 km/s, while some ($\sim$10\% to 25\%) have a
radial component less than 50 km/s. For the four BLAPs with spectral
observations, three of them have a radial-velocity difference larger
than 50 km/s and one less than 50 km/s, consistent with the above
distributions. We therefore conclude that our model well reproduces
the difference of the radial velocity between the BLAPs and the disc
stars, including the distribution of the difference, key evidence in
support of the surviving companion origin for BLAPs.

However, it must be emphasized that the current positions of the
BLAPs in the Galaxy are not their birth sites if they are the
surviving companions of SNe Ia. Based on the orbital velocity at
the time of the supernova explosion and the time passed since the
supernova, we estimate that they could travel $\sim2\,$kpc in the
Galaxy, which adds an additional uncertainty of the radial
velocity difference of as much as $20 - 40\,$km/s. Therefore, the
radial velocity difference between the BLAPs and the normal disc
stars in Fig.~\ref{rvdis} could be underestimated or overestimated
by as much as $20 - 40\,$km/s. Most importantly, a significant
radial velocity difference between the BLAPs and the normal disc
stars clearly exists, and the distribution of the radial component
of the orbital velocity in Fig.~\ref{rvdis} almost certainly can
explain the radial-velocity difference.

\begin{figure}
\centerline{\includegraphics[angle=270,scale=.35]{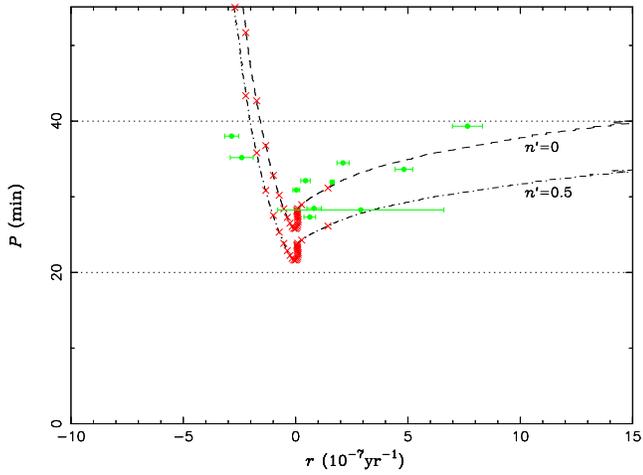}}
\caption{The evolution of the period and the rate of the period
change for the model with [$M_{\rm WD}^{\rm i}/M_{\odot}$, $M_{\rm
2}^{\rm i}/M_{\odot}$, $\log(P^{\rm i}/{\rm d})$] = (1.1, 3.0,
0.8) for different $n'$. The age interval between adjacent
crosses is $10^{\rm 6}$ yr. The green points show 11 BLAPs from
\citet{PIETRUKOWICZ17}, while the two dotted lines show the period
range for the whole BLAP's sample.}\label{osrmin}
\end{figure}

\subsection{The pulsation period}\label{sect:2.5} BLAPs are mysterious
variables, and, at present, it is completely unclear which
mechanism drives their pulsations. Two processes could play an
important role, one is the metal opacity bump at
$T\simeq2\times10^{\rm 5}~{\rm K}$ and the second is radiative
levitation of iron (\citealt{PIETRUKOWICZ17}). In
\citet{PIETRUKOWICZ17}, it is difficult to identify the pulsation
mode of BLAPs based on their light curve alone, although a radial
fundamental mode pulsation is favored (see also
\citealt{MCWHIRTER20}). \citet{PIETRUKOWICZ17} presented the
measured pulsation period and the rate of the period change. In
our model, there is only a very thin convective zone when the
companion star crosses the region of BLAPs in the HR diagram.
Therefore their pulsation modes cannot be simple, solar-type
oscillations. Here, we estimate the characteristic p-mode
frequency using

\begin{equation} \nu_{n,
l}\approx(n+\frac{l}{2}+\epsilon)\Delta\nu,\label{nu}
\end{equation}
which is independent of the detailed driving
mechanism (\citealt{BROWN91}; \citealt{KJELDSE95}). When both
radial order $n$ and angular degree $l$ are 0, we may obtain the
frequency of the fundamental mode by

 \begin{equation}
 \nu_{\rm 0}\approx\epsilon\Delta\nu=134.6\epsilon\frac{(M/M_{\odot})^{1/2}}{(R/R_{\odot})^{3/2}}~{\rm
\mu Hz},\label{f0}
  \end{equation}
where $\Delta\nu$ is the mean large-frequency separation of a star
and is calibrated to the Sun (\citealt{KJELDSE95};
\citealt{YANGWM09}), and $\epsilon$ is a constant and set to
$2.6$ for BLAPs (\citealt{WUT18}). Then, we can estimate the
pulsation period as
 \begin{equation}
 P=\frac{10^{\rm 6}}{60}\frac{1}{(n'+\epsilon)\Delta\nu}~{\rm min},\label{p0}
  \end{equation}
where $n'=n+\frac{l}{2}=0,~ 1/2,~1,~3/2,~2, \ldots$.
Following the definition of the rate of the period change in
\citet{PIETRUKOWICZ17}, we define the rate as
 \begin{equation}
 r=\frac{\Delta P}{\Delta t}\frac{1}{P}=\frac{P_{\rm i+1}-P_{\rm i}}{t_{\rm i+1}-t_{\rm i}}\frac{1}{P_{\rm i+1}}.\label{rate}
  \end{equation}
Varying $n'$ to fit the observational data, we find that our model
could reproduce the observations for $n'=0$ or $n'=0.5$, as
shown in Fig.~\ref{osrmin}. However, it is difficult to arrive at a
definitive conclusion on the oscillation mode based on the results
presented here. We cannot clearly distinguish between the radial
fundamental mode or a non-radial p-mode oscillation. The observations
of the color index of BLAPs seem to favour radial fundamental
pulsations (\citealt{PIETRUKOWICZ17}). Interestingly, $n'=0$ is the
radial fundamental mode. In addition, Equation~\ref{f0} shows that
$\nu_{\rm 0}$ is determined by the mean density of the star,
i.e.\ stars with similar masses and similar radii will have similar
$\nu_{\rm 0}$. This is the reason why BLAPs have similar surface
gravities and are locate in similar regions in the HR diagram.
Fig.~\ref{blap} shows that most surviving companions have a mass
around $0.76~M_{\odot}\pm0.1~M_{\odot}$. When these stars cross the
region of BLAPs in the HR diagram, they will also have similar radii
and similar surface gravities.

In addition, based on the results here, there is an evolutionary
sequence for BLAPs with a negative and a positive rate of period
change, i.e.\ before stars have reached the lowest luminosity in the
BLAP region, they show a negative rate of period change, while they
have a positive rate of period change thereafter (see also
Fig.~\ref{hrd}). Based on the models in \citet{WUT18}, the sign of the
rate of period change reflects the central helium abundance of the
star, $Y_{\rm c}$, i.e.\ stars with $Y_{\rm c}>0.45$ show a negative
rate of period change, while those with $Y_{\rm c}<0.45$ have a
positive rate of period change, consistent with our results. It is
worth emphasizing that, besides having a shorter lifetime than the
observed BLAPs, the shell-hydrogen-burning model can only explain
BLAPs with a negative rate of period change (\citealt{BYRNE18};
\citealt{CORSICO18}; \citealt{WUT18}), while our model can produce
BLAPs with both negative and positive rates simultaneously. This again
suggests that BLAPs are in the middle or late core-helium-burning
phase, as shown in Fig.~\ref{hrd} (see also
\citealt{WUT18}). Therefore, our model naturally explains why BLAPs
have similar surface gravities, similar pulsation periods and similar
rates of period change, including their sign, at the same time. These
results are not very surprising as it has previously been shown that
stars in the middle or late core-helium-burning phase with a mass of
$\sim0.7-1.1$ $M_{\odot}$ can reproduce the pulsation properties of
BLAPs (\citealt{WUT18}).

\begin{figure}
\centerline{\includegraphics[angle=270,scale=.35]{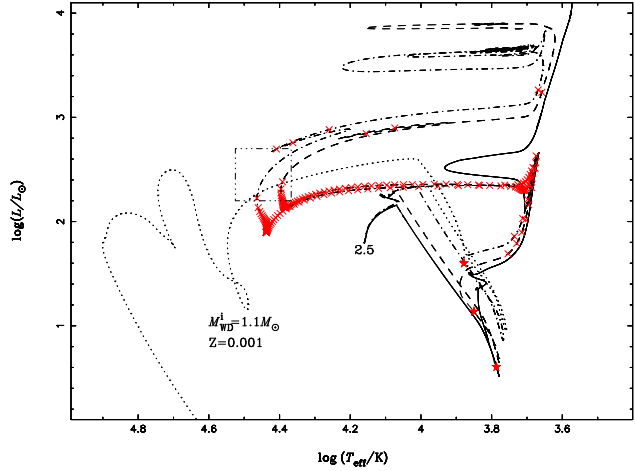}}
\caption{The evolutionary tacks of the companions for systems with
$Z=0.001$ and ($M_{\rm WD}^{\rm i}/M_{\odot}$, $M_{\rm 2}^{\rm
i}/M_{\odot}$) = (1.1, 2.5) for different initial periods, i.e.\
$\log(P^{\rm i}/{\rm d})$=0.0, 0.1, 0.3 and 0.4, respectively. The red
stars show the positions where the supernova explosions are assumed to
occur. The age interval between adjacent crosses is $10^{\rm 6}$
yr. The dash-triple-dotted rectangle indicates the region of
BLAPs.}\label{hrd001} \end{figure}

\section{DISCUSSION}\label{sect:3}
\subsection{Metallicity}\label{sect:3.1} In
this paper, we propose that BLAPs are the surviving companions of SNe
Ia. Our model can naturally reproduce all the properties of BLAPs,
including their single-star nature, the lifetime as a BLAP, their
position in the HR diagram, the surface helium abundance and gravity,
the total number of BLAPs in the Galaxy, the distribution of their
radial velocities, the pulsation periods, and the rate of the period
change, including the sign of the change. In addition, BLAPs are
relatively young objects and are not discovered in low-metallicity
environments (\citealt{PIETRUKOWICZ18}). This is also a natural
consequence of the metallicity dependence on the initial parameter
space for SNe Ia.  The initial parameter space in $P^{\rm i}-M_{\rm
  2}^{\rm i}$ plane for BLAPs puts them in the right-upper region for
SNe Ia (see Fig.~\ref{gblap}); this means that the progenitor systems
of BLAPs must contain a relatively massive companion with a relatively
long orbital period. With a decrease of metallicity, it becomes more
difficult for systems located in this region of parameter space to
become SNe Ia. For $Z\leq0.001$, no system in this region produces a
SN Ia because of violent nova explosions preventing an increase of the
mass of the WDs (see Fig.\ 4 in \citealt{MENG09}). Therefore the model
predicts that BLAPs cannot be produced in a very low-metallicity
environment, naturally explaining why BLAPs favor a young population
with relatively high metallicity.

While this is based on the OTW model, we also did several binary
evolution calculations with $Z=0.001$ to test whether the above
discussion still holds for our CEW model. Based on the results in
\citet{MENGXC17a}, we estimate that the upper boundary of the
companion mass in the $P^{\rm i}-M_{\rm 2}^{\rm i}$ plane for
$Z=0.001$ from our CEW model would be higher than that from the OTW
model. We use systems with ($M_{\rm WD}^{\rm i}/M_{\odot}$, $M_{\rm
  2}^{\rm i}/M_{\odot}$) = (1.1, 2.5) but different initial periods to
test the upper-right boundary in the initial $P^{\rm i}-M_{\rm 2}^{\rm
  i}$ plane, where the initial companion mass of 2.5 $M_{\odot}$ is
larger than the upper-boundary mass of the initial parameter space for
SNe Ia from the OTW model. The evolutionary tracks of these companions
in the HR diagram are shown in Fig.~\ref{hrd001}. As expected, the
upper-right boundary for SNe Ia from the CEW model is higher than that
from the OTW model by about 0.1$\,M_{\odot}$, which indicates that the
birth rate of SNe Ia from the CEW model is higher than that from the
OTW model\footnote{The OTW does not work when $Z$ is lower than a
  certain value (e.g.\ $Z<0.002$, \citealt{KOB98}), but the results in
  \citet{MENG09} are based on the assumption that the OTW is still
  valid for low metallicities. If the metallicity constraint is
  considered (e.g.\ $Z=0.001$), the upper-right boundary of the
  parameter space for SNe Ia and the birth rate of SNe Ia from the CEW
  model are probably significantly higher than those from the OTW
  model.} (\citealt{MENGXC17a}). However, some evolutionary tracks
still cross the region of BLAPs in the HR diagram, but their lifetimes
in the BLAP stage are short, and hence they are much less likely to be
found as BLAPs than for $Z=0.02$. In addition, for a star with a
similar mass at the same evolutionary stage, a low metallicity implies
a lower radius (\citealt{UMEDA99}; \citealt{CHENXF07};
\citealt{MENGXC08}), which is the main reason why the companions with
$Z=0.001$ spend most of their lives below the region of BLAPs in the
HR diagram. The lower radius indicates that the surviving companions
with $Z=0.001$ could not show the pulsations of BLAPs (see
Eq.~\ref{f0}) for most of their lifetimes. Moreover, generally, the
luminosity of the companions with $Z=0.001$ in the helium-core-burning
phase is lower than that of BLAPs, which indicates that a more massive
helium core and hence a larger initial orbital period would be
required to produce a BLAP. However, a system with a larger initial
period, even if it is only larger by 0.1 dex, will evolve to a system
of a WD + sdB star (the dotted line in Fig.~\ref{hrd001}) rather than
a SN Ia because of violent nova explosions and hence not produce a
BLAP.

The surviving companions of SNe Ia may be polluted by some heavy
elements, in particular Ni and Fe, as supernova ejecta pass the
companion (\citealt{MAR00}; \citealt{MENGXC07};
\citealt{PAKMOR08}). Therefore, the surface abundance of such
heavy elements on the surviving companions could be higher than
that for typical disc stars. Enhanced heavy elements could also be
helpful in producing the observed BLAP pulsations because of the
increased heavy-element opacity (\citealt{PIETRUKOWICZ17};
\citealt{ROMERO18}). However, before the companions become BLAPs,
a large convective region is likely develop in the envelope, mixing such
heavy elements from the supernova ejecta into the interior and making
such anomalies unobservable. At the same time, at high effective
temperature, radiative levitation effects could bring the inner
iron-group elements to the surface of the star in the BLAP stage,
as observed in the spectra of some subdwarf O stars
(\citealt{CHAYER95}; \citealt{CHARPINET97}; \citealt{LATOUR17}).
Generally, the timescales for the above two effects are much
shorter than the typical evolutionary timescale of hot subdwarf
stars (\citealt{DORMAN93}; \citealt{CHARPINET97}); hence BLAPs are
likely to have lost the information on the chemical pattern due to
any pollution by supernova ejecta. In any case, currently, with
only moderate-resolution spectra available, the abundance for the
heavy elements cannot be determined. Even if higher abundances
were to be determined by future observations, this would not
constitute key evidence in support of the surviving companion
nature of BLAPs.

If the initial metallicity of the progenitor system were not to affect
the production of BLAPs, we would expect about $10-100$ BLAPs in the
Large Magellanic Cloud (LMC), based on the birth rate of BLAPs in the
Galaxy and the star formation history in the LMC
(\citealt{HARRIS09}). However, no BLAPs have so far been observed in
the LMC and SMC (\citealt{PIETRUKOWICZ18}). This indicates that the
initial metallicity plays an important role in the production of
BLAPs, probably by affecting the parameter space for SNe Ia, as
discussed above based on the SD model for SNe Ia. In addition,
radiative levitation is probably required to produce the pulsation
modes of BLAPs, and an enhancement of iron and nickel could be a key
factor in the development of the pulsations seen in BLAPs
(\citealt{JEFFERY16}; \citealt{ROMERO18}; \citealt{BYRNE18}). Our
surviving companion scenario for BLAPs provides a natural explanation
for the enhancement of iron and nickel by the pollution from supernova
ejecta.

\subsection{Uncertainties}\label{sect:3.2}
In this paper, we assumed that a CO WD explodes as a SN Ia when
$M_{\rm WD}=1.378~M_{\odot}$.  We then followed the evolution
of the companion star and found that some companions can reproduce
the properties of BLAPs. However, there are two other effects which
could influence the companions and change the initial parameter
space producing BLAPs. One is the collision of supernova
ejecta with the companions (\citealt{MAR00}; \citealt{MENGXC07};
\citealt{PAKMOR08}; \citealt{LIUZW12}), the other is the so-called
spin-up/spin-down model (\citealt{JUSTHAM11};
\citealt{DISTEFANO12}).

For the SD model, the supernova ejecta may collide with the
envelope of the companion and strip off part of the envelope. The
amount of material stripped off is heavily dependent on the
structure of the companion (\citealt{MENGXC07};
\citealt{PAKMOR08}). For the systems leading to BLAPs, the
collision by the supernova ejecta may strip off about
$0.065\,M_{\odot}$ to $0.125\,M_{\odot}$ from the surface of the
companion, mainly depending on the ratio of binary separation to
the companion radius at the moment of supernova explosion
(\citealt{MENGXC07}; \citealt{PANK12}; \citealt{LIUZW12}). The
stripped hydrogen-rich material may reveal itself by narrow
H$\alpha$ emission line in their late-time spectrum
(\citealt{MAR00}; \citealt{MENGXC07}). However, such prediction
was not confirmed by the observations to most of SNe Ia
(\citealt{MAGUIRE16}; \citealt{TUCKER20}). On the other side, the
narrow H$\alpha$ emission line was indeed detected in some SNe Ia,
but the amount of the hydrogen-rich material deduced from
observations is much smaller than the theoretical predictions
(\citealt{MAGUIRE16}; \citealt{PRIETO20}). At present, the reason
of the confliction between observations and theories is still
unclear.

After the impact, the companion may be heated and expand quickly
to a luminosity as high as a few $10^{\rm 3}$ $L_{\odot}$, and the
deposited energy in the envelope of the companion will take a
thermal timescale to release (\citealt{MAR00};
\citealt{PODSIADLOWSKI03}; \citealt{SHAPPEE13}; \citealt{PANK14}).
Then, during this period, the companion may introduce an extra
stellar wind, reducing the envelope mass further. However,
assuming a simple Reimer's wind and taking a typical value of the
luminosity ($10^{\rm 3}$ $L_{\odot}$), the radius ($10^{\rm 2}$
$R_{\odot}$) and the thermal time scale ($10^{4}$ yr) of the
companion (\citealt{SHAPPEE13}), the companion would lose about
$10^{-4}$ $M_{\odot}$ during this period. Therefore, such effect
can be neglected.

The WDs in the systems may spin up as they gain angular momentum from the
accreted material. Rapidly rotating WDs, however, may exceed the
classical Chandrasekhar mass limit, and rotating super-Chandrasekhar WDs
must require a spin-down phase before they can explode as SNe Ia
(\citealt{JUSTHAM11}; \citealt{DISTEFANO12}). The
spin-down timescale is currently quite uncertain, probably between $10^{\rm
5}$ yr and $10^{\rm 7}$ yr (\citealt{DISTEFANO11};
\citealt{MENGXC13}). Based on the CE mass and the mass-loss rate
at the moment when $M_{\rm WD}=1.378~M_{\odot}$, \citet{MENGXC17b}
estimated that a spin-down timescale of $\sim10^{\rm 6}$ yr is
favoured. During the spin-down phase, the companion may continue
to lose envelope material. However, since the companion only has a very
thin envelope, the mass-lose rate would decrease quickly
to less than $10^{\rm -7}~M_{\odot}/{\rm yr}$, even stopping completely
(this is the main reason why the OTW model cannot produce SNe Ia
in the upper-right region of the initial parameter space, while
the CEW can; see Fig.~\ref{compare} and the detailed
discussions in \citealt{MENG09}), the companion may not lose too
much material during the spin-down phase, i.e.\ probably less than
0.1 $M_{\odot}$.

Therefore, the effects discussed above on the mass of the
companion are similar, i.e.\ decrease the companion mass at the
time of the supernova explosion. This could make the companions
more similar to hot subdwarf stars rather than BLAPs when helium
is ignited in the center, as in the model with [$M_{\rm WD}^{\rm
i}$, $M_{\rm 2}^{\rm i}/M_{\odot}$, $\log(P^{\rm i}/{\rm d})]$ =
(1.1, 3.3, 0.9) shows. In this case, models like [$M_{\rm WD}^{\rm
i}$, $M_{\rm 2}^{\rm i}/M_{\odot}$, $\log(P^{\rm i}/{\rm d})]$ =
(1.1, 3.0, 0.6) would become the progenitors of BLAPs. Therefore,
the main consequence of these effects would be to change the
initial parameter space producing BLAPs, i.e. the initial
parameter space moves to shorter initial period in
Fig.~\ref{gblap}. Hence, the predicted number of BLAPs here could
be underestimated or overestimated (see Fig. 11 in
\citealt{MENG09}). However, at present, the number of BLAPs in our
Galaxy is quite uncertain, and even if the uncertainty of the
theoretical predicted number is as high as 100\%, the number of
BLAPs predicted here is still consistent with the present
observational constraint. Therefore, the effects discussed above
would not significantly affect our main conclusions.

Besides the influence on the companion mass, these two effects
might also change the space velocity of the surviving companions
but in different directions. Compared with the orbital velocity
when $M_{\rm WD}=1.378~M_{\odot}$, the collision of supernova
ejecta on the companion increases its space velocity due to a kick
velocity imparted, but the kick velocity would be significantly
smaller than the orbital velocity.  (\citealt{MAR00};
\citealt{MENGXC07}; \citealt{MENGXC17a}). In contrast, a spin-down
phase may significantly decrease the orbital velocity. For
example, if a spin-down timescale of a few $10^{\rm 6}\,$yr is
considered, the orbital velocity of the companion at the time of
the supernova explosion would be in the range of $50-190\,$km/s
(\citealt{MENGXC18}). Therefore, the effect of the spin-down
mechanism is likely to dominate in determining the final space
velocity of the surviving companions. On both the observational
and the theoretical side, a spin-down phase seems likely to be
necessary (\citealt{SOKER17}; \citealt{MENGXC17b}), which means a
smaller space velocity for the surviving companions of SNe Ia than
that shown in Fig.~\ref{blap}. Therefore, the proportion of
systems with radial velocity less than $50\,$km/s in
Fig.~\ref{rvdis} is likely to be underestimated.

The GAIA project provides a unique opportunity to constrain the origin of
BLAPs. However, for the 14 BLAPs in \citet{PIETRUKOWICZ17},
only BLAP-009 has a reliable parallax measurement in GAIA DR2.
Based on the proper motion and distance from GAIA DR2 data and the
radial velocity in Fig.~\ref{rvblap}, we can obtain
the components of the space velocity of BLAP-009 in the Milky
Way's Galactic coordinate system: $U=39.8\pm35.1$ km/s,
$V=182.9\pm13.4$ km/s, and $W=13.4\pm2.5$ km/s
(\citealt{ASTRAATMADJA16}; \citealt{LURI18}). Hence, BLAP-009 has a
lower space velocity than a typical disc star around the location of
BLAP-009 by $\sim60\pm22$ km/s, i.e.\ BLAP-009 may even be taken as
a runaway star, considering the large difference of the space velocity
(\citealt{BLAAUW61}; \citealt{BROWNWR15}; \citealt{HUANGY16}).
The velocity difference of $60\pm22$ km/s is smaller than
the prediction in Fig.~\ref{blap}, but is consistent with the results
in \citet{MENGXC18}, which would imply that a spin-down phase is
necessary for the production of BLAPs if they are the
surviving companions of SNe Ia\footnote{Although the distance of
other BLAPs are not as precise as for BLAP-009 in GAIA DR2, we show
the components of the space velocity of BLAP-014 in the Milky
Way's Galactic coordinate system as a reference, whose distance is
relatively precise compared to other BLAPs, i.e.\ $U=35.6\pm34.9$
km/s, $V=159.6\pm61.4$ km/s, and $W=-11.9\pm11.7$ km/s. Therefore,
BLAP-014 has a lower space velocity than a typical disc star
around the location of BLAP-014 by $\sim79\pm62$ km/s.}.

The measurement of the masses of BLAPs could provide a key clue to
constrain the origin of BLAPs; but unfortunately there are
too many uncertainties for estimating the masses of the BLAPs
due to uncertainties in the distances, brightnesses, surface
gravities and effective temperatures. If all these uncertainties are
considered, the mass of the BLAP-009 could be anywhere between 0.06 $M_{\odot}$
and 1.40 $M_{\odot}$, hence not providing a meaningful constraint.

\begin{figure}
\centerline{\includegraphics[angle=270,scale=.35]{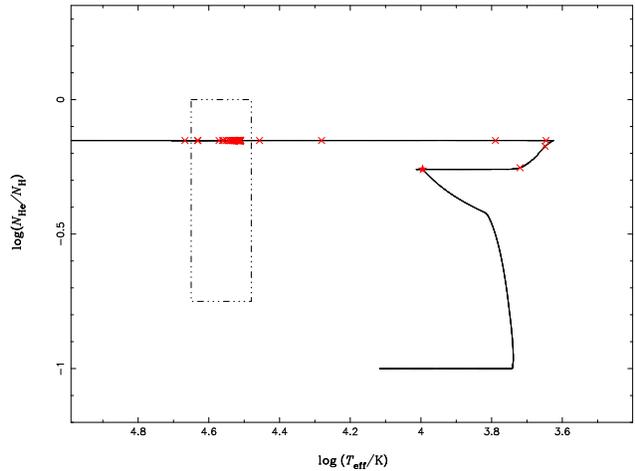}}
\caption{The evolution of the surface helium abundance and effective
temperature for the model with [$M_{\rm WD}^{\rm i}/M_{\odot}$,
$M_{\rm 2}^{\rm i}/M_{\odot}$, $\log(P^{\rm i}/{\rm d})$] = (1.1, 3.3,
0.9). The dash-triple-dotted rectangle presents the region for group 4
of hot sundwarf stars in \citet{LUOYP16} (private communication). Red
stars show the position where supernova explosions are assumed to
occur, and the age interval between adjacent crosses is $10^{\rm 6}$
yr.}\label{txhe} \end{figure}

\subsection{Other possible
origins}\label{sect:3.3} As discussed in \citet{PIETRUKOWICZ17},
BLAPs are core-helium-burning or hydrogen-shell-burning stars with
thin envelopes. For the hydrogen-shell-burning model, BLAPs are
possibly the progenitors of extremely low-mass WDs with high
effective temperatures and a stellar mass of
$\sim0.34~M_{\odot}$\footnote{Recently, \citet{KUPFER19} found a
new class of BLAPs with higher surface gravities and proposed that
this new class of BLAPs are extremely low-mass WDs. We will
address this class in another paper.} (\citealt{ROMERO18}). But
the lifetime of the hydrogen-shell-burning stars in the BLAP stage
is too short to be compatibile with the long-term stability of the
pulsation periods of observed BLAPs with a typical timescale of
$10^{\rm 7}$ yr (\citealt{PODSIADLOWSKI02}; \citealt{WUT18}). Even
if the progenitors of extremely low-mass WDs were to contribute to
the BLAP population, they could only produce BLAPs with a negative
rate of period change (\citealt{BYRNE18}; \citealt{CORSICO18};
\citealt{WUT18}). Therefore, the core-helium-burning model is the
favoured model for BLAPs, as predicted by our model. Nevertheless,
there are potentially several other channels to form such a
structure. However, as discussed in the following, no other
channel currently considered can explain all the properties of
BLAPs simultaneously, and every alternative channel has its
problems. As we will show now, only the surviving companion
scenario may be able to solve all problems simultaneously.

To form the structure of a BLAP, a star needs to lose its envelope
in the HG or on the first giant branch (FGB). Since BLAPs are
single stars, single-star channels need to be considered. For a
single star with $M_{\rm i}\leq1.0\,M_{\odot}$ and metallicity
$Z\geq0.02$, a star may lose most or all of its envelope near the
tip of the FGB because these envelopes are extremely weakly bound
(\citealt{HAN94}; \citealt{MENGXC08}). If a thin envelope remains
and helium is ignited in the center of the remnant after envelope
ejection, the star would show the main properties of BLAPs. Such a
channel could easily explain the dependence of the BLAPs on
metallicity. However, BLAPs from such a channel would belong to an
old population, which is inconsistent with the young population
nature of the BLAPs. In addition, this origin would not explain
the unusual radial velocity of BLAPs. If this channel contributes
to BLAPs, there should be many BLAPs in old metal-rich clusters,
e.g. NGC 6791, but no BLAPs have been reported in NGC 6791. Also,
if helium is ignited in the center after envelope ejection, the
star is more likely to become a hot subdwarf star rather than a
BLAP (\citealt{KALIRAI07}; \citealt{STEINFADT12};
\citealt{HANZW13}).

Another possible channel to form the structure of a BLAP is also from
the SD scenario for SNe Ia, where the companion of the WD is a
red-giant (RG) star, i.e.\ comes from the WD + RG channel. After the
supernova explosion, the supernova ejecta may strip off almost all the
envelope of the RG companion (\citealt{MAR00}). If the hydrogen shell
is still burning, the companion could show the properties of BLAPs
(e.g.\ the shell-hydrogen-burning model in
\citealt{PIETRUKOWICZ17}). However, this channel also has problems
with the population and radial velocity as discussed above
(\citealt{WANGB10}). In addition, the envelope of the companion
after the collision with the supernova ejecta is so thin (i.e.\ less
than 0.02 $M_{\odot}$) that the lifetime of the companion in the
shell-burning stage is too short (i.e.\ shorter than $10^{\rm 5}$ yr)
to explain the long-term stability of the pulsation periods of BLAPs
with a typical timescale of $10^{\rm 7}$ yr.  After the extinction of
the hydrogen shell, helium generally cannot be ignited in the center
of the star because of its low mass, and the companion probably becomes
a low-mass single WD rather than a BLAP (\citealt{JUSTHAM09};
\citealt{MENGYANG10}; \citealt{MENGXC16}).  Even if helium were
ignited in the center, the companion would appear as a hot subdwarf
with a mass of less than 0.45 $M_{\odot}$ rather than a BLAP
(\citealt{MENGXC13}).

\citet{PIETRUKOWICZ17} discussed a possible origin from the
Galactic Center, i.e.\ the progenitors of BLAPs would be members
of binary systems passing the central supermassive black hole,
where the companions are captured by the supermassive black hole
while the progenitors of the BLAPs are ejected from the Galactic
Center. However, the position of BLAPs in the Galaxy and their
radial velocities do not support such a runaway scenario. The
conclusive evidence to exclude the runaway scenario comes from the
GAIA observation for BLAP-009, as discussed in section
\ref{sect:3.2}. The components of the space velocity of BLAP-009
in the Milky Way's Galactic coordinate system and its distance of
$5.50\pm0.53$ kpc to the Galactic Centre clearly prove that it
cannot originate from the Galactic Centre.

Since BLAPs could be related to hot subdwarfs, another channel to form
single hot-subdwarf stars could also contribute to BLAPs, i.e.\ if the
progenitor of a BLAP is a FGB star in a binary system. If the
companion of the FGB is a low-mass star or a brown dwarf, possibly
even as small as a planet, the system could merge during a CE phase
and form a rapidly rotating HB star. The centrifugal force for rapid
rotation may enhance the mass loss from the surface of the HB star and
a BLAP might form (\citealt{SOKER98}; \citealt{POLITANO08}).  This
scenario could easily explain why BLAPs seem to be connected with hot
subdwarf stars, but it is difficult to explain the distribution of
their radial velocities and their young-population nature. Similarly,
the merger of two helium WDs to form a single hot subdwarf star also
does not explain the metallicity dependence and the unusual radial
velocities. Moreover, the merger scenario of two helium WDs is
expected to produce extremely hydrogen-deficient hot subdwarf stars,
inconsistent with BLAPs (\citealt{ZHANGXF12}).

\citet{PIETRUKOWICZ17} could not exclude the possibility that some
BLAPs have very faint companions. So, subdwarf
stars in long-period binary systems could contribute to the BLAP population
(\citealt{HANZW02,HANZW03}; \citealt{CHENXC13}). However, such a
channel has the same problems as the previous models with the
metallicity dependence and radial velocity distribution of BLAPs.

There is another puzzle for channels related to the formation of
normal hot subdwarf stars: why have BLAPs only been discovered
recently in contrast to hot subdwarf stars. The most reasonable
explanation is that the formation process for BLAPs is not
associated with the normal hot-subdwarf channel, and that the
number of BLAPs is much smaller than that of normal hot subdwarf
stars. Based on the results in this paper and \citet{HANZW03}, we
may estimate that the theoretical number ratio of single hot
subdwarf stars to BLAPs lies roughly between 6 and 640. Currently,
about 2000 hot subdwarf stars have been confirmed
spectroscopically, but the single-star frequency among them is
still uncertain (\citealt{GEIER15,GEIER17};
\citealt{KEPLER15,KEPLER16} \citealt{LUOYP16}). It probably lies
between 10\% and 50\% (see the discussion in \citealt{HANZW03}).
So, the number of discovered single hot subdwarfs lies roughly
between 200 and 1000; this would imply that the observational
number ratio of single hot subdwarf stars to BLAPs is between 14
and 71, consistent with our theoretical estimates. When the total
catalogue of hot subdwarf stars before the GAIA mission is
considered, the ratio may increase up to 200, still in the range
of the theoretical estimates (\citealt{GEIER17}). Therefore, the
theoretical and the observed number ratios appear consistent with
each other, at least at the present observational level.

Our model makes a prediction on the distribution of BLAPs in
the Galaxy. The progenitors are born in the thin disc and then spread
in all directions. Since there is continuous star formation in
the thin disc, we may expect that the number density of BLAPs in
the thin disc is the highest, with a lower value in the thick
disc, and the lowest in the halo. Future surveys may be able to check
this prediction.

\subsection{A new channel to form single hot subdwarf stars}\label{sect:3.4}
As we showed in this paper, some systems from
the same channel that produces BLAPs but with slightly different
initial parameters can produce single hot subdwarf stars (see
Fig.~\ref{hrd}). This is in fact a new channel to form single hot
subdwarf stars which may have different properties from those forming
from other, more canonical evolutionary scenarios, e.g.\ (a) the
merger of two helium WDs, (b) the merger of a FGB star and its
low-mass companion and (c) the envelope ejection scenario for single
low-mass high-metallicity FGB stars (see the discussions in the above
section and \citealt{HANZW02,HANZW03}; \citealt{HEBER09};
\citealt{HAN94}; \citealt{MENGXC08}).

1) Generally, hot subdwarf stars from scenario a) are extremely
helium-rich sdOs with strong N lines in their atmospheres
(\citealt{HEBER09}; \citealt{ZHANGXF12}). These extremely helium-rich
sdOs usually have a $\log(n_{\rm He}/n_{\rm H})$ larger than 0, even
larger than 1 (\citealt{LUOYP16}), while the single hot sub\-dwarf stars
from scenarios (b) and (c) usually have a $\log(n_{\rm He}/n_{\rm H})$
less than $-1$. However, the single hot subdwarf stars from our model
generally have a medium $\log(n_{\rm He}/n_{\rm H})$ value,
i.e.\ between 0 and $-1$ (see Fig.~\ref{gxhe}).

2) The mass of the single subdwarf stars from scenarios (a), (b) and
(c) has a broad range from 0.3 $M_{\odot}$ to 0.8$\,M_{\odot}$ and
peaks at the canonical mass for the He core-flash at 0.46 $M_{\odot}$
(\citealt{HANZW03}; \citealt{MENGXC08}; \citealt{POLITANO08};
\citealt{HANZW13}), while the single hot subdwarf stars from our model
have a mass larger than 0.5 $M_{\odot}$, up to 0.97 $M_{\odot}$
(\citealt{MENGXC17a}).

3) The present single hot-subdwarf-star sample is mainly discovered in
the thick disc or halo of the Galaxy, which means that they belong to
a relatively old population (\citealt{LUOYP16}), while the hot
subdwarf stars from our model belong to a young population and could
be discovered in the thin or thick disc of the Galaxy.

4) Compared to scenarios (a), (b) and (c), the hot subdwarf stars from
our model inherit the orbital velocities of the binary systems at the
time of the supernova explosion and will show a different space
velocity. In addition, Figs~\ref{compare} and \ref{gblap} show that
the OTW model has difficulties in producing such single hot subdwarf
stars. Hence, the discovery of such hot subdwarf stars will favour our
CEW model.

Interestingly, there exists a small group in the current
hot-subdwarf-star sample, consistent with our predictions but
difficult to be explained by standard binary evolutionary channels
(e.g.\ group 4 in Fig.~8 of \citealt{LUOYP16}).  Fig.~\ref{txhe}
shows the evolution of the surface helium abundance and the effective
temperature of the companions from the systems with [$M_{\rm WD}^{\rm
i}/M_{\odot}$, $M_{\rm 2}^{\rm i}/M_{\odot}$, $\log(P^{\rm i}/{\rm
d})$] = (1.1, 3.3, 0.9). The figure shows that the companion after the
supernova explosion spends most of its life in the region of group 4
in \citet{LUOYP16}; hence our model provides a reasonable origin for
this group. Also, the figure shows that, if the spin-down timescale is
as long as 6 Myr, the companion star could become a hot subdwarf star
before the supernova explosion (\citealt{MENGXC18}).  Such a spin-down
timescale of a rapidly rotating WD is consistent with the estimate
in \citet{MENGXC13}. This result could open a new window for
searching for a surviving companion in a supernova remnant or the progenitor
system in archival images taken before the supernova explosion
(\citealt{MENGXC18}). However, the properties of the hot subdwarf
stars from our SN Ia channel could be difficult to distinguish from
those originating from the CE merger channel, except that the
atmosphere of the hot subdwarf stars from the SNe Ia channel could be
polluted by supernova ejecta. However, the heavy elements from
supernova ejecta pollution would not be a good tracer to distinguish
different origins, as discussed in section~\ref{sect:3.1}. One
possible mechanism to distinguish the hot subdwarf stars from these
two channels is to measure the radial velocity since the radial
velocity of the stars from the SN Ia channel is generally larger than
that from other channels.  Moreover, the distribution of such single
hot subdwarf stars in the Galaxy provides another clue to distinguish them
from other single hot subdwarf stars since they are mainly located in the
thin disc and few should be found in the halo, similar to
BLAPs. We will investigate this channel in more detail in the future.\\

\section{Conclusion}\label{sect:4}
In summary, we propose that the mysterious BLAPs are the surviving
companions of SNe Ia, since all the properties of the BLAPs may be
reasonably reproduced by our SD model simultaneously, including
their population characteristics, positions in the HR diagram,
spectroscopic properties, radial velocities, pulsation periods and
their rates of period change and the total number of BLAPs in the
Galaxy. No other proposed channel can simultaneously explain all
these properties. We predict the distribution of BLAPs in the
Galaxy, with their number density being highest in the thin disc,
lower in the thick disc and lowest in the halo. We also predict a
new channel for {\em single} hot subdwarf stars, which connects
them directly to BLAPs with a generally high radial velocity. Such
single hot subdwarf stars have a similar space distribution as
BLAPs in the Galaxy. If such hot subdwarf stars are confirmed
observationally, this would provide additional support for our CEW
model. Interestingly, there already exists a small group of
objects in the currently known single hot-subdwarf-star sample
with properties consistent with our model predictions.

\section*{Acknowledgments}
We are grateful to Pawel Pietrukowicz, Yan Li and Tao Wu for their
kind helps and discussions. This work was supported by the NSFC
(No. 11973080, 11521303 and 11733008). X.M. acknowledges the
support by the Yunnan Ten Thousand Talents Plan¡ªYoung \& Elite
Talents Project.

\end{document}